\documentclass[aps, showpacs, preprint]{revtex4-1}
\usepackage{amssymb, graphicx, latexsym, amsfonts, bm}

\begin{document}

\title{Non-modal analysis of the diocotron instability. Plane geometry}
\author{V.V. Mikhailenko}
\altaffiliation{Pusan National University, 30 Jangjeon-dong, Guemjeong-gu, Pusan 609--735, S. Korea.}
\email[E-mail: ]{vladimir@pusan.ac.kr}
\author{H.J. Lee}%
\affiliation{Pusan National University, 30 Jangjeon-dong, Guemjeong-gu, Pusan 609--735, S. Korea.}
\author{V.S. Mikhailenko}
\affiliation{V.N. Karazin Kharkov National University, 61108 Kharkov, Ukraine.}

\date{\today}

\begin{abstract}
The comprehensive investigation of the temporal 
evolution of the diocotron instability of the plane electron strip on the linear 
stage of its development is performed. By using the Kelvin's method of the 
shearing modes we elucidate the role of the initial perturbations of the 
electron density, which is connected with problem of the continuous 
spectrum. The linear non-modal evolution process, detected by the solution 
of the initial value problem, leads towards convergence to the phase-locking 
configuration of the mutually growing normal modes.
\end{abstract}

\pacs{52.27.Gr}
\keywords{diocotron instability; non-neutral plasma}
\maketitle




\section{Introduction}

The diocotron instability is one of the most ubiquitous instabilities\cite{Davidson} in  
low-density non-neutral plasmas with shear in the flow velocity.
The goal of this study is to gain further insight into the physics of the 
development of the diocotron instability, assessing the role of the continuous 
spectrum \cite{Briggs, Case} and the recognition the dynamics of the relative 
phase of the diocotron waves on the linear stage of the instability development.
These two problems are of paramount importance in understanding 
the temporal evolution of the initial perturbations into patterns peculiar 
to diocotron instability and has the major practical motivations. The 
solution of both these problems is amenable to examination using an 
initial value formulation of the linearised diocotron instability problem.

Since the first papers\cite{Levy65} devoted to the theory of the diocotron 
instability, the complete analogy of the stability problem of two-dimensional cold low-density 
electron beams in crossed external magnetic and self electric fields, 
and the problem of stability of the incompressible inviscid shear flows was recognized. 
Last time, valuable progress was 
obtained in the investigations of the stability of shear flows in a geophysical 
context\cite{Heifetz, Heifetz2004, Bakas}. A future common to these 
investigations is the recognition that a key element in the understanding 
the dynamics of the instability development, formation of the edges pattern and development of the 
vortex structures is the relative phase of the edge waves. In this paper 
we adopt this view point to exemain inter-level interaction of edge diocotron 
waves in what is probably the simplest non-trivial case
-- we consider the plane model of the unbounded along the coordinate $y$ 
non-neutral electron flow, which is confined in the strip $-a\leq x\leq a$
and moves with ExB velocity in homogeneous magnetic field 
$\mathbf{B}=B\mathbf{e}_{z}$ and in own crossed electric field, 
$\mathbf{E}=E\left(x \right)\mathbf{e}_{x}$, of the electrons strip.
Our analysis is grounded on the method of the shearing modes (or so-called non-modal approach),
developed in first by Lord Kelvin\cite{Kelvin} (see, also Refs.\cite{Mikhailenko, Mikhailenko-2005}).
In the next Section, we formulate the basic non-modal 
equations. In Sec.III, these equations are solved for the case, 
when the initial perturbations of the electron density may be ignored and give the modal theory 
of the diocotron instability. In Sec.IV, we consider the diocotron instability 
in terms of edge waves interactions. The non-modal analysis of the diocotron 
instability, is presented in Sec.V, in which the role of the initial perturbations
of the electron density on the temporal evolution of the diocotron instability 
to the phase-locking configuration, discovered in Sec.IV, is analysed. 
A summary of the work is given in Conclusions, Section VI.

\section{Basic equation}
The basic equation in the theory of the diocotron instability in plane geometry is
the equation for the perturbed electrostatic potential $\phi$ \cite{Davidson}
\begin{eqnarray}
&\displaystyle
\left(\frac{\partial}{\partial t}+V'_{0}x\frac{\partial}{\partial y}\right)\nabla^{2}\phi\left(x,y,t \right) =
\frac{4\pi ec}{B_{0}}\frac{\partial\phi}{\partial y}\frac{dn_{0}}{dx}.  \label{1}
\end{eqnarray}
In this paper we consider the homogeneous basic density of electrons, for 
which velocity shearing rate is $V'_{0y}=\omega^{2}_{pe}/\omega_{ce}$, 
where $\omega_{pe}$ and $\omega_{ce}$ are electron plasma frequency and electron 
cyclotron frequency, respectively.
We consider  tenuous electron layer satisfying $\omega_{pe}\ll\omega_{ce}$,  
confined in a strip with edges at $x=\pm a$, i.e.  $n_{0}\left(x\right)= 
n_{0}\left(\Theta\left(x+a\right)-\Theta\left(x-a\right) \right)$, 
where $\Theta\left(x-a\right)$ is the unit-step Heaviside function 
(it is equal to zero for $x<a$ and equal to unity for $x\geq a$); thus,
$dn_{0}\left(x\right) /dx
= n_{0}\left(\delta\left(x+a\right)-\delta\left(x-a\right) \right)$. 
In this case Eq.(\ref{1}) is written in the form
\begin{eqnarray}
&\displaystyle
\left( \frac{\partial}{\partial t}+V'_{0}x\frac{\partial}{\partial y}\right)
\nabla^{2}\phi\left(x,y,t \right) 
\nonumber\\&\displaystyle
=V'_{0}\left(\delta\left(x+a\right)-\delta\left(x-a\right) \right)
\frac{\partial\phi}{\partial y}.  \label{2}
\end{eqnarray}
Now we define boundary conditions for potential $\phi\left(x, y, t\right)$. 
We suppose, that potential is continuous across the surfaces $x=\pm a$, 
i.e. $\phi\left(x=\pm a - \epsilon, y,t\right) = 
\phi\left(x=\pm a + \epsilon, y,t\right)$
with $\epsilon \rightarrow 0$. The conditions on the 
jump of the $d\phi/dx$ at $x=\pm a$ 
are determined by the integration\cite{Levy65} of Eq.(\ref{1}) 
for the short distances $\pm\epsilon\rightarrow 0$ 
across the both surfaces $x=\pm a$,
\begin{eqnarray}
&\displaystyle
\left( \frac{\partial}{\partial t}+V'_{0}a\frac{\partial}
{\partial y}\right)\left[\left.\frac{\partial\phi}{\partial x}\right|_{x=a+\epsilon} 
- \left.\frac{\partial\phi}{\partial x}\right|_{x=a-\epsilon}  \right] 
\nonumber\\&\displaystyle
=-V'_{0}\frac{\partial}{\partial y}\phi\left(x= a , y,t\right)
\label{3}
\\ &\displaystyle
\left( \frac{\partial}{\partial t}-V'_{0}a\frac{\partial}{\partial y}\right)
\left[\left.\frac{\partial\phi}{\partial x}\right|_{x=-a+\epsilon} 
- \left.\frac{\partial\phi}{\partial x}\right|_{x=-a-\epsilon}  \right] 
\nonumber\\&\displaystyle
=V'_{0}\frac{\partial}{\partial y}\phi\left(x= -a , y,t\right) \label{4} 
\end{eqnarray}
Also we require, that potential decays in vacuum region, i.e. $\phi
\left(x=\pm\infty, y, t\right) = 0$.

We describe two areas: the electron layer, $-a\leqslant x\leqslant a$, 
and vacuum in the rest of space. Eq.(\ref{2}) in the vacuum  have a form
\begin{eqnarray}
&\displaystyle
\frac{\partial}{\partial t}\nabla^{2}\phi =0. \label{5} 
\end{eqnarray}
The Fourier transformed over $y$ solutions $\phi\left(x, l,t\right)=
\int \phi\left(x, y,t\right)\exp\left(-ily\right) dy$ of Eq.(\ref{5}) are
\begin{eqnarray}
&\displaystyle
\phi\left(x, l,t\right)=C_{1}\left(l,t\right)e^{-lx} \qquad \text{for}
 \qquad x>a,\label{6}
\end{eqnarray}
\begin{eqnarray}
&\displaystyle
\phi\left(x, l, t\right)=C_{2}\left(l,t\right)e^{lx} \qquad \text{for}
 \qquad x<-a.  \label{7}
\end{eqnarray}
In electron layer, the right hand side side of Eq. (\ref{2}) 
is equal to zero, except the edges at $x=\pm a$, i.e.
\begin{eqnarray}
&\displaystyle \left( \frac{\partial}{\partial t}+V'_{0}x
\frac{\partial}{\partial y}\right)\nabla^{2}\phi =0 .  \label{8}
\end{eqnarray}
In the usually applied normal-mode (modal) approach, the solution to Eq.(\ref{8})
is sought in the form $\phi=\phi\left(x \right)\exp\left(-i\omega t+ik_{y}y\right)$, 
for which Eq.(\ref{8}) transforms to the form\cite{Levy65, Knauer}
\begin{eqnarray}
&\displaystyle \left(\omega-k_{y}V'_{0}x\right)\nabla^{2}\phi 
=0 .  \label{9}
\end{eqnarray}
It is assumed, that $\omega-k_{y}V'_{0}x\neq 0$, and equation $\nabla^{2}\phi
=0$ is solved\cite{Levy65, Knauer}. The proper accounting for 
the resonance $\omega-k_{y}V'_{0}x=0$ is attained by solving the 
initial value problem to Eq.(\ref{9})\cite{Case, Briggs}. 
The principal result of the solution of the initial value problem 
by using the Laplace transform in time\cite{Case, Briggs} was that, in addition to the 
discrete eigenvalues linked to the normal modes, 
there exists a continuous spectrum of eigenvalues. 
The application of the Laplace transform to the solution of that problem\cite{Case, Briggs} made 
the calculation complicate and low effective - the explicit results for the 
temporal evolution of the potential were obtained only for the asymptotically large time.

Here we use other approach, which gives easy and transparent
treating of the problem considered without the application of the spectral transforms in time. 
It is known, that for homogeneous shear flows, the solution of the  initial value problem
is greatly facilitated by a transformation of coordinates $x,y$ to 
a new set of coordinate frame $\xi, \eta$, that is sheared with the 
mean flow. Such a transformation was proposed by Lord Kelvin\cite{Kelvin} and is given
by (see also Refs.\cite{Mikhailenko, Mikhailenko-2005})
\begin{eqnarray}
&\displaystyle
x=\xi, \qquad y=\eta+V'_{0}\xi t, \qquad t=t.  \label{10}
\end{eqnarray}
In the convected coordinate frame, Eq.(\ref{8}) is spatially homogeneous 
and the inhomogeneity introduced by flow velocity in Eq.(\ref{8}) 
is transformed to a temporal inhomogeneity,
\begin{eqnarray}
&\displaystyle
\frac{\partial}{\partial t}\left[\frac{\partial^{2}\phi }
{\partial \xi^{2}} -2 V'_{0}t\:\frac{\partial^{2}\phi }
{\partial \xi\partial \eta}
+\left(1+\left(V'_{0} t \right)^{2}  \right) 
\frac{\partial^{2}\phi }{\partial\eta^{2}}\right]
=0.  \label{11}
\end{eqnarray}
The integration of Eq.(\ref{11}) over time, 
\begin{eqnarray}
&\displaystyle
\frac{\partial^{2}\phi}{\partial \xi^{2}} -2 V'_{0}t\:
\frac{\partial^{2}\phi}{\partial \xi\partial \eta}
+\left(1+\left(V'_{0} t \right)^{2}  \right)  \frac{\partial^{2}\phi }{\partial\eta^{2}}
\nonumber\\&\displaystyle
= 4\pi e n_{1}\left(t=0,\xi,\eta \right),  \label{12}
\end{eqnarray}
brings into the considered problem the initial 
perturbation of the density of electrons in layer, $n_{1}\left(t=0,\xi,\eta \right)$. 
After Fourier transforming over $\eta$ of Eq. (\ref{12}), we have
\begin{eqnarray}
&\displaystyle
\frac{\partial^{2}\phi}{\partial \xi^{2}} -2i V'_{0}lt\:
\frac{\partial\phi }{\partial \xi}
-l^{2}\left(1+\left(V'_{0} t \right)^{2}  \right)\phi
\nonumber\\&\displaystyle
= 4\pi e n_{1}\left(t=0,\xi,l \right),  \label{13}
\end{eqnarray}
The solution to Eq.(\ref{13}) for potential $\phi_{1}$ straightforwardly 
gives the initial value problem solution for the separate spatial 
Fourier harmonic for any desired time (without usual application\cite{Case}
of the Laplace transform with time). That solution, 
\begin{eqnarray}
&\displaystyle
\phi\left(\xi,l,t \right) =\left( C_{3}\left(l,t \right) + 
\frac{2\pi e}{l}\int\limits^{\xi}_{-a}d\xi_{1}n_{1}\left(\xi_{1},l\right
)e^{-k_{1}\xi_{1}}\right) e^{k_{1}\xi} 
\nonumber \\&\displaystyle +\left(C_{4}\left(l,t \right)
+\frac{2\pi e}{l}\int\limits^{a}_{\xi}d\xi_{1}n_{1}\left(\xi_{1},l\right)
e^{-k_{2}\xi_{1}}\right)e^{k_{2}\xi},  \label{14} 
\end{eqnarray}
has an obvious non-modal form with non-separable dependences on 
time and coordinate in exponentials, resulted from the 
time-dependent $k_{1,2}=\pm l + iV'_{0}lt$. 
The temporal evolution of the potential
appears to be the strictly non-modal process.

\section{Modal diocotron instability} 

If we suppose that any initial perturbation in layer 
are absent, i.e. $n_{1}\left(\xi_{1}, l \right)=0$, then
we obtain the solution, which describes only the surface 
waves, which form the discrete spectrum of perturbations
\begin{eqnarray}
&\displaystyle
\phi\left(\xi,l \right)= C_{3}\left(l,t \right)e^{k_{1}\xi}
+C_{4}\left(l,t \right)e^{k_{2}\xi}.  \label{15}
\end{eqnarray}

The condition of the continuity of the perturbed potential 
on the boundaries ($x=\xi=\pm a$) couples $C_{1}\left(l,t \right)$, 
$C_{2}\left(l,t \right)$ and $C_{3}\left(l,t \right)$, 
$C_{4}\left(l,t \right)$, and gives the following presentation for the potential:
\begin{eqnarray}
&\displaystyle
\phi\left(\xi,l \right)= \Big( C_{3}\left(l,t \right)
+C_{4}\left(l,t \right)e^{2al}\Big) e^{l\xi-iV'_{0}alt}, \xi<-a
\nonumber\\&\displaystyle
\phi\left(\xi,l \right)= \Big( C_{3}\left(l,t \right
)e^{l\xi}+C_{4}\left(l,t \right)e^{-l\xi}\Big)
e^{iV'_{0}\xi lt}, \quad -a<\xi<a  ,  \label{16}
\\&\displaystyle
\phi\left(\xi,l \right)= \Big( C_{3}\left(l,t \right)e^{2al}
+C_{4}\left(l,t \right)\Big) e^{-l\xi+iV'_{0}alt}, \xi>a.
\nonumber
\end{eqnarray}

Now we apply boundary conditions (\ref{3}), (\ref{4}) to (\ref{16}), 
and obtain the system of equations for $C_{3}\left(l,t \right)$ 
and $C_{4}\left(l,t \right)$, i.e.
\begin{eqnarray}
&\displaystyle
\frac{\partial C_{3}}{\partial t}=i\frac{V'_{0}}{2}\left(1-2la \right)C_{3}
+i\frac{V'_{0}}{2}e^{-2al}C_{4},  \label{17}
\\&\displaystyle
\frac{\partial C_{3}}{\partial t}=-i\frac{V'_{0}}{2}e^{-2al}C_{3}
-i\frac{V'_{0}}{2}\left(1-2la \right)C_{4}. \label{18}
\end{eqnarray}

\begin{figure}
\includegraphics[width=0.4\textwidth]{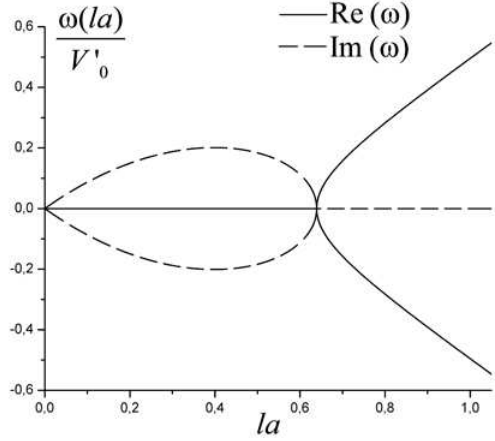}
\caption{\label{fig1}The dispersion relation of the normal modes.}
\end{figure}

The solution to Eqs. (\ref{17}), (\ref{18}) gives the relation
\begin{eqnarray}
&\displaystyle
\omega =\pm \frac{V'_{0}}{2}\sqrt{\left(1-2la \right)^{2}-e^{-4al}}.  \label{19}
\end{eqnarray}
which is the known dispersion equation for diocotron oscillations in plane 
charge sheet\cite{Knauer}. The dispersion equation (\ref{19}) is illustrated 
in Fig.1. as a function of the parameter $al$.
In the case $e^{-4al}>\left(1-2la \right)^{2}$, we have imaginary 
frequency which defines the growth rate of the diocotron instability. 
In the case, when boundaries of electron layer are so far from each 
other that $e^{-4al}<\left(1-2la \right)^{2}$, we have two not interacting stable waves. 

\section{Modal diocotron instability interpreted in terms of edge waves interaction}

The result of the above analysis is the known dispersion equation (\ref{19}). 
This equation, however, does not give an overview of how different initial 
wave structures will evolve in time. Writing the functions $C_{3}\left(l,t \right)$ 
and $C_{4}\left(l,t \right)$ in complex form\cite{Heifetz}, 
\begin{eqnarray}
&\displaystyle
C_{3}\left(l,t \right)= Q_{3}\left(l,t \right)e^{i\epsilon_{3}\left(l,t \right)},\label{20}
\\&\displaystyle
C_{4}\left(l,t \right)= Q_{4}\left(l,t \right)e^{i\epsilon_{4}\left(l,t \right)}, \label{21}
\end{eqnarray}
the edge perturbation of the potential can be regarded as two edge waves 
with amplitudes $ Q_{3}\left(l,t \right)$ and $ Q_{4}\left(l,t \right)$ 
and phases $\epsilon_{3}\left(l,t \right)$ and $\epsilon_{3}\left(l,t \right)$. 
By substituting Eqs.(\ref{20}), (\ref{21}) into Eqs.(\ref{17}), (\ref{18}) and 
separating the real and imaginary parts at $x=\pm a$, we obtain, that 
amplitudes $Q_{3}\left(l,t \right)$ and $Q_{4}\left(l,t \right)$, and 
the relative phase $\epsilon=\epsilon_{3}-\epsilon_{4}$ of the edge 
diocotron waves evolve according to equations 
\begin{eqnarray}
&\displaystyle
\frac{d Q_{3}}{d t}=\frac{V'_{0}}{2}e^{-2al}Q_{4} \sin\epsilon, \label{22}
\\
&\displaystyle
\frac{d Q_{4}}{d t}=\frac{V'_{0}}{2}e^{-2al}Q_{3} \sin\epsilon, \label{23}
\\
&\displaystyle
\frac{d \epsilon}{d t}=\Gamma\left(\cos\epsilon+b\left(t \right)\right),  \label{24}
\end{eqnarray}
where
\begin{eqnarray}
&\displaystyle
\Gamma=\frac{V'_{0}}{2}e^{-2al}\left( \frac{Q_{3}}{Q_{4}}+\frac{Q_{4}}{Q_{3}}\right),  \label{25}
\end{eqnarray}
and 
\begin{eqnarray}
&\displaystyle
b\left(t\right)=\frac{2\left( 1-2la\right)e^{2al}}{\left(\frac{Q_{3}}{Q_{4}}
+\frac{Q_{4}}{Q_{3}}\right)},  \label{26}
\end{eqnarray}
It follows from  (\ref{19}), that at the condition 
of the diocotron instability development, $ \left(1-2la\right)e^{2al}<1$. 
Also, it is easily obtained, that
\begin{eqnarray}
&\displaystyle
\frac{Q_{3}}{Q_{4}}+\frac{Q_{4}}{Q_{3}}\geq 2  \label{27}
\end{eqnarray}
(the equality sign appears when $Q_{3}=Q_{4}$). From Eqs.(\ref{22}), 
(\ref{23}) one can obtain the integral,
\begin{eqnarray}
&\displaystyle
Q^{2}_{3}-Q^{2}_{4}=C.  \label{28}
\end{eqnarray}
Therefore, due to the exponential growth of amplitudes $Q_{3}$, $Q_{4}$ 
with time from infinitesimal beginnings, the amplitudes become almost 
equal, $Q^{2}_{3}=Q^{2}_{4}+C\approx Q^{2}_{4}\gg C$, and $b\left(t\right)$ 
approaches the value $b_{0}=\left( 1-2la\right)e^{2al}$.
Therefore,  at the condition, under which the diocotron instability 
develops, we have $b\left(t\right)<1$, and therefore,  the stationary 
(or fixed) points\cite{Strogatz} of the equation (\ref{24}), where 
$d\epsilon/dt=0$, exist and are determined by the equation 
$\cos\epsilon+b_{0}=0$. The solutions of this 
equation are two sets of stationary points: stable (or attractors) at
\begin{eqnarray}
&\displaystyle
\epsilon_{k}=\left(\pi-\cos^{-1} b_{0}\right)+2k\pi, \label{29}
\end{eqnarray}
and unstable at 
\begin{eqnarray}
&\displaystyle
\epsilon_{k}=-\left(\pi-\cos^{-1} b_{0}\right)+2k\pi. \label{30}
\end{eqnarray}
Note, that solution of the equation $d\epsilon/dt=\cos \epsilon +b_{0}$ 
with initial condition $\epsilon=\epsilon_{0}$ at $t=t_{0}=0$, has a simple form
\begin{eqnarray}
&\displaystyle
\tan \frac{\epsilon}{2}=-\sqrt{\frac{1+b_{0}}{1-b_{0}}}
\left(\frac{1+Ae^{t\sqrt{1-b^{2}_{0}}}}{1-Ae^{t\sqrt{1-b^{2}_{0}}}}\right) \label{31}
\end{eqnarray}
where
\begin{eqnarray}
&\displaystyle
A=\frac{\left(1-b_{0} \right) \tan \frac{\epsilon_{0}}{2}+\sqrt{1-b^{2}_{0}}}
{\left(1-b_{0} \right) \tan \frac{\epsilon_{0}}{2}-\sqrt{1-b^{2}_{0}}}\label{32}
\end{eqnarray} 
As it follows from Eq.(\ref{31}), the initial perturbations with arbitrary value of 
the initial phase of each wave, will evolve with time to the ultimate value of relative phase,
\begin{eqnarray}
&\displaystyle\tan \frac{\epsilon}{2}=\sqrt{\frac{1+b_{0}}{1-b_{0}}}
\label{33}
\end{eqnarray}
which does not depend on the initial data and is the same as
established by Eq.(\ref{29}) (see Fig.2). For $al=0.4$, Eq.(\ref{33}) gives $\epsilon\approx 110^{o}$
(see Fig.3, where that phase locking configuration is presented). 
This solution of the initial value problem reveals 
the initial stage of the instability development as a process of the formation of the phase 
locked configuration. As it follows from Eqs.(\ref{22})--(\ref{24}), the time 
of the developing of the phase locking configuration is comparable 
with of  the inverse growth rate time, $t\gtrsim \Gamma^{-1}\simeq\left( V'_{0}\right)^{-1}$ 
of the diocotron instability. Two edge diocotron waves, 
embedded within spatially distinct regions of oppositely directed 
density gradients interact such, that the wave trains transit to 
a phase-locked state of mutual growth.

\begin{figure}
\includegraphics[width=0.4\textwidth]{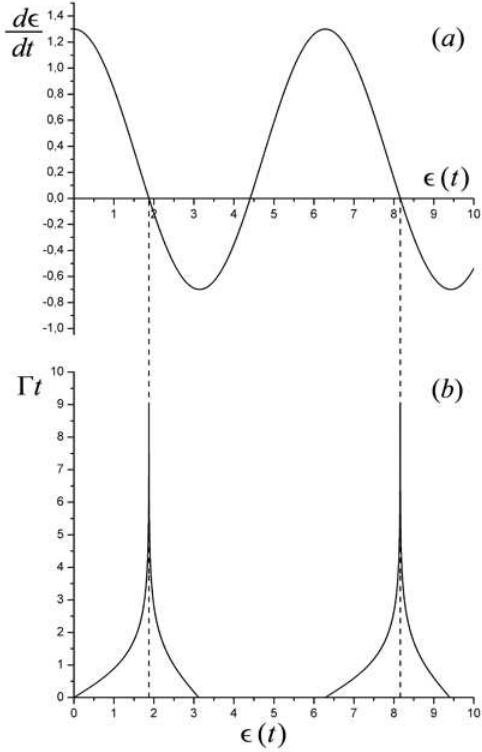}
\caption{\label{fig2} (a)Phase portrait for Eq.(24); (b) The evolution 
of the relative phase $\epsilon$ with time.}
\end{figure}

\section{Non-Modal analysis of the diocotron instability.}

In the considered above idealized case, when the initial perturbation $n_{1}\left(t=0, \xi,\eta\right)$ of the 
electron density is ignored, Eqs. (\ref{17}), (\ref{18}) resulted from the boundary 
conditions at $\xi=\pm a$, in which non-modal non-separable structure as $\xi t$ 
of the spatio-temporal dependence
of the solution (\ref{14}) receives ordinary modal, as $at$, form. As a result, 
we obtain modal theory of the diocotron instability. The accounting for 
the initial perturbation of the electron density excludes such simplifications 
and complete non-modal analysis becomes necessary. In this section, we solve 
for solution (\ref{14})
the boundary value problem, determined by the condition of the continuity of 
the potential $\phi_{1}$ and by Eqs.(\ref{3}), (\ref{4}). The continuity of 
the potential gives the following presentation of the solutions (\ref{6}), (\ref{7}) through the 
functions $C_{3}\left(l,t \right) $ and $C_{4}\left(l,t \right)$ of the solution (\ref{14}) in different 
regions of the space. We obtain for $\xi<-a$
\begin{figure}
\includegraphics[width=0.4\textwidth]{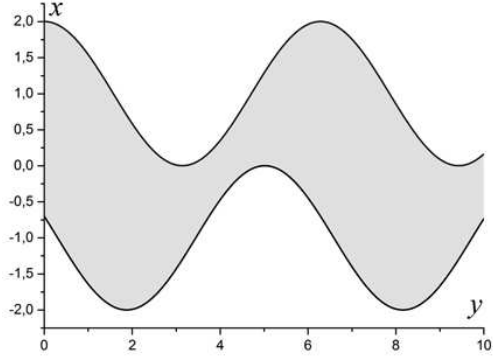}
\caption{\label{fig3}The phase-locking configuration with relative phase $\epsilon\approx 110^{o}$, 
obtained from Eq.(\ref{30}) for $al=0.4$.}
\end{figure}
\begin{eqnarray}
&\displaystyle
\phi\left(\xi,l,t \right) =\Big(C_{3}\left(l,t \right)e^{-k_{1}a} 
+C_{4}\left(l,t \right)e^{-k_{2}a}\nonumber \\&\displaystyle\left.    
+\frac{2\pi e}{l}\int\limits^{a}_{-a}d\xi_{1}n_{1}\left(\xi_{1},l \right
)e^{-k_{1}\left(\xi_{1}+a \right)}\right)
e^{l\left(\xi +a \right)},  \label{34} 
\end{eqnarray}
and for $\xi>a$
\begin{eqnarray}
&\displaystyle\phi\left(\xi,l,t \right) =\Big(C_{3}\left(l,t \right)
e^{k_{1}a} +C_{4}\left(l,t \right)e^{k_{2}a} \nonumber \\&\displaystyle\left.    
+\frac{2\pi e}{l}\int\limits^{a}_{-a}  d\xi_{1}n_{1}\left(\xi_{1},l \right)
e^{-k_{1}\left(\xi_{1}-a \right)}\right)e^{-l\left(\xi -a \right)},  \label{35} 
\end{eqnarray}
The application of the conditions (\ref{3}), (\ref{4}) to  solutions (\ref{14}), 
(\ref{34}), and (\ref{35})
gives the inhomogeneous equations for $C_{3}\left(l,t \right) $ and $C_{4}\left(l,t \right)$, 
\begin{eqnarray}
&\displaystyle\frac{\partial C_{3}}{\partial t}=i\frac{V'_{0}}{2}
\left(1-2la \right)C_{3}\left(l,t \right)
+i\frac{V'_{0}}{2}e^{-2al}C_{4}\left(l,t \right)+f_{1}\left(l,t \right),  \label{36}
\\&\displaystyle\frac{\partial C_{3}}{\partial t}=-i\frac{V'_{0}}{2}
e^{-2al}C_{3}\left(l,t \right)
-i\frac{V'_{0}}{2}\left(1-2la \right)C_{4}\left(l,t \right)+f_{2}\left(l,t \right). \label{37}
\end{eqnarray}
where
\begin{eqnarray}
&\displaystyle
f_{1}\left(l,t \right)= i\pi e\frac{V'_{0}}{l}\int\limits^{a}_{-a}d\xi_{1}n_{1}\left(\xi_{1},l \right)e^{-\left(l+iV'_{0}t \right) \xi_{1} }
\left(1+2l\left( \xi_{1}-a\right)  \right),  \label{38} 
\end{eqnarray}
\begin{eqnarray}
&\displaystyle
f_{2}\left(l,t \right)= -i\pi e\frac{V'_{0}}{l}\int\limits^{a}_{-a}d\xi_{1}n_{1}
\left(\xi_{1},l \right)e^{\left(l-iV'_{0}t \right) \xi_{1} }
\left(1-2l\left( \xi_{1}+a\right)  \right).  \label{39} 
\end{eqnarray}
The solution to system (\ref{36})-(\ref{39}) for $C_{3}\left(l, t \right)$ and 
$C_{4}\left(l, t \right)$ is obtained  straightforwardly and is given  by 
\begin{eqnarray}
&\displaystyle
C_{3}\left(l, t \right)=c_{1}e^{\gamma t}+c_{2}e^{-\gamma t}+ \hat{C_{3}}\left(l, t \right),  \label{40} 
\end{eqnarray}
and 
\begin{eqnarray}
&\displaystyle
C_{4}\left(l, t \right)=c_{1}\alpha_{1}e^{\gamma t}+c_{2}\alpha_{2}e^{-\gamma t}
+ \hat{C_{4}}\left(l, t \right),  \label{41} 
\end{eqnarray}
where 
\begin{eqnarray}
&\displaystyle
\hat{C_{3}}\left(l, t \right)=\frac{\pi e \left(V'_{0} \right)^{2} }{4l\gamma}e^{-2la}\sum_{m=-\infty}^{\infty}n_{1}\left(m,l \right)e^{im\pi} 
\nonumber \\&\displaystyle\times\left(e^{\gamma t}\left(-\alpha_{2}I_{1}+ I_{2}\right)+e^{-\gamma t}\left(-I_{3}+\alpha_{1}I_{4}\right)\right), \label{42}
\\
&\displaystyle
\hat{C_{4}}\left(l, t \right)=\frac{\pi e \left(V'_{0} \right)^{2} }{4l\gamma}e^{-2la}\sum_{m=-\infty}^{\infty}n_{1}\left(m,l \right)e^{im\pi} 
\nonumber \\&\displaystyle\times\left(\alpha_{1}e^{\gamma t}\left(-\alpha_{2}I_{1}+I_{2}\right)
+\alpha_{2}e^{-\gamma t}\left(\alpha_{1}I_{3}-I_{4}\right)\right).\label{43}
\end{eqnarray}
In Eqs.(\ref{42}), (\ref{43})
\begin{eqnarray}
&\displaystyle I_{1,3}=\int\limits^{t}dt_{1}\frac{e^{-k_{1}a\mp\gamma t_{1}}}{i\frac{m\pi}{a}-k_{1}}
\left(1-\frac{2l}{i\frac{m\pi}{a}-k_{1}}\right) \nonumber \\&\displaystyle -\int\limits^{t}dt_{1}\frac{e^{k_{1}a\mp\gamma t_{1}}}{i\frac{m\pi}{a}-k_{1}}
\left(1-4al-\frac{2l}{i\frac{m\pi}{a}-k_{1}}\right),\label{44}
\end{eqnarray}
\begin{eqnarray}
&\displaystyle I_{2,4}=\int\limits^{t}dt_{1}\frac{e^{k_{2}a\mp\gamma t_{1}}}{i\frac{m\pi}{a}-k_{2}}
\left(1-\frac{2l}{i\frac{m\pi}{a}-k_{2}}\right) \nonumber \\&\displaystyle +\int\limits^{t}dt_{1}\frac{e^{-k_{2}a\mp\gamma t_{1}}}{i\frac{m\pi}{a}-k_{2}}
\left(-1+4al-\frac{2l}{i\frac{m\pi}{a}-k_{2}}\right), \label{45}
\end{eqnarray}
where $k_{1,2}=\pm l + iV'_{0}lt$,  $\gamma=i\omega$ is the growth rate, 
determined by Eq.(\ref{19}), $\alpha_{1,2}=-e^{2la}\left(\left(1-2la \right)\pm\frac{2i\gamma}{V'_{0}}\right)$,
and the presentation of the initial density perturbation in a form of Fourier series,
\begin{eqnarray*}
&\displaystyle
n_{1}\left(\xi,l \right)=\sum_{m=-\infty}^{\infty}n_{1}\left(m,l \right)
e^{\frac{im\pi\xi}{a}},
\end{eqnarray*} 
was used. Eqs.(\ref{14}), (\ref{34}), (\ref{35}), and (\ref{42})--(\ref{45}) 
compose the complete explicit solution on the initial and boundary value problems 
for the separate Fourier harmonic $\phi\left(\xi,l, t \right)$. That solution 
is valid for all times, at which linear theory of the diocotron instability is applicable. 
As it follows from Eqs.(\ref{44}), (\ref{45}), at time $t<t_{*}=\pi m/laV'_{0}$ the 
denominators in integrands of $I_{1,3}$ and $I_{2,4}$ decrease with time and approach 
their minimal value $\pm l$ at time $t=t_{*}$, which is of the order of a few inverse 
growth rate times. Only after time $t_{*}$, denominators grow with 
time as $t$, leading to the decay of the potential with time as $\left
(V'_{0}t\right)^{-1}$. It is important 
to note, that only that last stage of the potential evolution in time is amenable analytically 
by using inverse Laplace transform\cite{Case, Briggs}, 
temporal growth of the potential at time $t<t_{*}$,  known as the Orr 
mechanism\cite{Orr, Bakas}, was completely overlooked.

The solution for $\phi\left(\xi,l, t \right)$ for time $t>t_{*}$ is easily obtained 
by integration of $I_{1,3}$ and $I_{2,4}$ by parts and may be presented in the form
\begin{eqnarray}
&\displaystyle
\phi\left(\xi,l, t \right)=\phi_{(0)}\left(\xi,l, t \right)+\hat{\phi}\left(\xi,l, t \right),
\label{46}
\end{eqnarray}
where
\begin{eqnarray}
&\displaystyle
\phi_{(0)}\left(\xi,l, t \right)=c_{1}\left( e^{\gamma t+k_{1}\xi}
+\alpha_{1}e^{\gamma t+k_{2}\xi}\right) \nonumber \\&\displaystyle
+c_{2}\left(e^{-\gamma t+k_{1}\xi}+\alpha_{2}e^{-\gamma t
+k_{2}\xi}\right).\label{47}
\end{eqnarray} 
The constants $c_{1}$ and $c_{2}$ are determined by the initial 
perturbations of the electrostatic potential (or electron density) 
on the boundary surfaces at $x=\pm a$. The potential $\hat{\phi}$ 
is equal to
\begin{eqnarray}
&\displaystyle
\hat{\phi}\left(\xi,l, t \right)= \hat{\phi}_{1}\left(\xi,l, t \right)
+\hat{\phi}_{2}\left(\xi,l, t \right)+\hat{\phi}_{3}\left(\xi,l, t \right),
\label{48}
\end{eqnarray}
where
\begin{eqnarray}
&\displaystyle
\hat{\phi}_{1}\left(\xi,l, t \right)=\hat{C_{3}}\left(l, t \right)e^{k_{1}\xi}  \nonumber 
\\&\displaystyle= -\frac{\pi e\left(V'_{0} \right)^{2} }
{4\gamma l}
e^{-2la}\sum_{m=-\infty}^{\infty}n_{1}\left(m,l \right)e^{im\pi} \nonumber 
\\&\displaystyle
\times\left\lbrace \frac{1}{k_{1}-\frac{i\pi m}{a}}\left[ e^{k_{1}\left(\xi-a \right) }
\left(\frac{\alpha_{2}}{\gamma+ilaV'_{0}}+\frac{\alpha_{1}}{\gamma-ilaV'_{0}} \right)
 \right.\right.\nonumber \\&\displaystyle\left.\left. -e^{k_{1}\left(\xi+a \right) }
 \left(1-4la \right) \left(\frac{\alpha_{2}}{\gamma-ilaV'_{0}}+\frac{\alpha_{1}}
 {\gamma+ilaV'_{0}} \right) \right]\right. \nonumber \\&\displaystyle\left.
 +\frac{2\gamma}{\left( \gamma^{2}+\left(V'_{0} la\right)^{2}\right)
 \left(\frac{i\pi m}{a}-k_{2} \right)}
\right. \nonumber \\&\displaystyle\left. \times\left(e^{k_{1}\left(\xi+a \right) }
e^{-2la}-\left(1-4a \right)e^{k_{1}\left(\xi-a \right)}e^{2la} \right) \right\rbrace, 
\label{49}
\end{eqnarray}

\begin{eqnarray}
&\displaystyle
\hat{\phi}_{2}\left(\xi,l, t \right)=\hat{C_{4}}\left(l, t \right)e^{k_{2}\xi}  \nonumber 
\\&\displaystyle =-\frac{\pi e\left(V'_{0} \right)^{2} }{4\gamma l}
e^{-2la}\sum_{m=-\infty}^{\infty}n_{1}\left(m,l \right)e^{im\pi} \nonumber 
\\&\displaystyle
\times\left\lbrace\frac{e^{k_{2}\left(\xi+a \right)}}{\left(\frac{i\pi m}{a}-k_{2}\right)}
\left(\frac{\alpha_{1}}{\gamma-ilaV'_{0}}+\frac{\alpha_{2}}
{\gamma+ilaV'_{0}} \right)\right.\nonumber \\&\displaystyle\left. 
-\frac{\left(1-4la \right)e^{k_{2}\left(\xi-a \right) }}{\left(\frac{i\pi m}
{a}-k_{2}\right)}\left(\frac{\alpha_{2}}{\gamma+ilaV'_{0}}+\frac{\alpha_{1}}
{\gamma-ilaV'_{0}} \right)\right. \nonumber \\&\displaystyle\left.
+\frac{2\gamma\alpha_{1}\alpha_{2}}{\left( \gamma^{2}+\left(V'_{0} la\right)^{2}\right)
\left(k_{1}-\frac{i\pi m}{a}\right)}
\right. \nonumber \\&\displaystyle\left. \times\left(e^{k_{2}\left(\xi-a \right) }
e^{-2la}-\left(1-4a \right)e^{k_{2}\left(\xi+a \right)}e^{2la} \right) \right\rbrace, 
\label{50}
\end{eqnarray}
and  $\hat{\phi}_{3}$ is formed by the initial perturbations in Eq.(\ref{14}),
\begin{eqnarray}
&\displaystyle
\hat{\phi}_{3}\left(\xi,l, t \right)=\frac{2\pi e}{l}
\sum_{m=-\infty}^{\infty}n_{1}\left(m,l \right) \nonumber \\&\displaystyle
\times\left[ \frac{ \left(e^{i\pi m}e^{k_{1}\left(\xi+a \right)}
-e^{\frac{i\pi m \xi}{a}} \right)}{\left(k_{1}-\frac{i\pi m}{a}\right)}+\frac{ \left(e^{i\pi m}e^{k_{2}\left(\xi-a \right)}
 -e^{\frac{i\pi m \xi}{a}}\right)}{\left(\frac{i\pi m}{a}-k_{2}\right)} \right]. 
\label{51}
\end{eqnarray}

The  presentation of the solution (\ref{46}) in coordinates $x, y, t$ is attained by 
performing the inverse Fourier transform for wavenumber $l$  with changing the conjugate variable 
$\eta$ using the transformations (\ref{10}),
\begin{eqnarray}
&\displaystyle
\phi\left(x, y, t \right)= \frac{1}{2\pi}\int dl e^{ily-iV'_{0}xtl}\left(\phi_{(0)}\left(x,l, t \right)\right.
\nonumber \\&\displaystyle\left.+\hat{\phi}\left(x,l, t \right)\right)=\phi_{(0)}\left(x, y, t \right)+
\hat{\phi}\left(x, y, t \right).\label{52}
\end{eqnarray} 
It is interesting to note, that $\phi_{(0)}\left(x, y, t \right)$ has a modal form,
\begin{eqnarray}
&\displaystyle
\phi_{(0)}\left(x, y, t \right)= \frac{1}{2\pi}\int dl e^{ily}\left(c_{1}\left(l\right)e^{\gamma t+lx}\right.
\nonumber \\&\displaystyle\left.+c_{2}\left(l\right)e^{-\gamma t+lx}+c_{1}\left(l\right)\alpha_{1}e^{\gamma t-lx}+c_{2}\left(l\right)\alpha_{2}e^{-\gamma t-lx} \right),\label{53}
\end{eqnarray} 
whereas $\hat{\phi}\left(x, y, t \right)$ is non-modal in both  sets of variables. 
The integrand, which determines $\hat{\phi}\left(x, y, t \right)$ in Eq.(\ref{52}), decays 
with time as $\left(V'_{0}t\right)^{-1}$ at times $t>t_{*}$ 
and contains non-modal multiplier $\exp\left(-iV'_{0}xtl\right)$ (which cancelled 
in Eq.(\ref{53})). From the spectral point of view,  solution (\ref{52}) reveals  the 
result, obtained in the asymptotic limit of large time by using Laplace transform\cite{Case, Briggs}, 
that the spectrum of the diocotron instability contains discrete spectrum 
of two modes (solution $\phi_{(0)}\left(x, y, t \right)$), and continuous spectrum, 
which is presented by $\hat{\phi}\left(x,l, t \right)$.  

\section{Conclusions}

In this paper, we have performed the comprehensive investigation of the temporal 
evolution of the diocotron instability of the plane electron strip on the linear 
stage of its development. 
In the realistic laboratory flows, as well as in numerical simulations, perturbation 
growth occurs over a finite time interval due to disruption by instabilities 
and turbulence. It is therefore of interest to find all factors yielding 
the instability development over a specified time interval.
We find, that normal mode theory, presented in Sec.III, is unable to 
explain the physics of the evolutionary processes, which occur in the volume 
and on surfaces of the electron strip. The understanding of these inherently 
non-modal processes attains via solution of the initial value problem. By using 
the Kelvin's method of shearing modes we elucidate the role of the 
initial perturbations of the electron density, which is connected with problem of the continuous 
spectrum \cite{Briggs, Case}.  Because the growth rate $\gamma\lesssim 0.2V'_{0}$,
at time $t\gtrsim \gamma^{-1}\backsim 5\left( V'_{0}\right) ^{-1}$, effect of the non-modality in the 
solution for $\phi\left(\xi,l, t \right)$ becomes important. We find, that linear 
non-modal evolution process, detected by the solution of the initial value problem, 
leads towards convergence to the phase-locking configuration of the growing normal modes, 
which is the precursor of the development of the multi vortex structure, 
observed in experiments and in numerical simulations.

It is important to note, that solutions (\ref{42})--(\ref{45}), which are valid for any times, 
incorporate additional mechanism of the temporal 
growth of the initial disturbances, known as the Orr mechanism\cite{Orr, Bakas}
 -- solutions (\ref{42})--(\ref{45}) experience the temporal growth at
$t\leq t_{*}$, at which  the denominators of Eqs.(\ref{44}), (\ref{45})
achieve their minimal values.  Depending on the magnitude 
of the initial perturbations of the electron density, this effect may be comparable 
with modal growth, and it is crucial for the growth of the short along the 
shear flow perturbations, which are stable against the modal diocotron instability.

\end{document}